\documentclass{aa}
\usepackage{graphicx}
\usepackage{txfonts}
\begin{document}
   \title{The AGN nature of three \emph{INTEGRAL} sources: IGR J18249--3243, IGR J19443+2117 and 
IGR J22292+6647\thanks{Partly based on observations collected at the Astronomical
Observatory of Bologna in Loiano (Italy).}}


    \titlerunning{The AGN nature of three \emph{INTEGRAL} sources}
\authorrunning{R.~Landi}

   \author{R. Landi\inst{1}, J.B. Stephen\inst{1}, N. Masetti\inst{1}, D. Grupe\inst{2},
F. Capitanio\inst{3}, A.J. Bird\inst{4}, A.J. Dean\inst{4}, M. Fiocchi\inst{3}, and  N. Gehrels\inst{5}
          }

   \offprints{landi@iasfbo.inaf.it}
\institute{INAF/IASF Bologna, via Piero Gobetti 101, I-40129 Bologna, Italy \and
Department of Astronomy and Astrophysics, Pennsylvania State University,
University Park, PA 16802 \and
INAF/IASF Roma, via Fosso del Cavaliere 100, I-00133 Roma, Italy \and
School of Physics and Astronomy, University of Southampton, Highfield, SO17 1BJ, UK \and
NASA Goddard Space Flight Center, Greenbelt, MD 20771
}
   \date{Received  / accepted}

  \abstract
   {The third \emph{INTEGRAL}/IBIS survey has revealed several new hard X-ray sources, which are still 
unidentified/unclassified. To identify these sources, we need to find their counterparts at other
wavelengths and then study their nature.}
   {The aim of this work is to employ the capability of the X-ray telescope (XRT) on board \emph{Swift} to 
localize the sources with a positional accuracy of few arcseconds, thus allowing the search for 
optical/UV, 
infrared and radio counterparts to be more efficient and reliable. A second objective is to provide 
spectral information over a broad range of frequencies.} 
  {We analysed all XRT observations available for three unidentified \emph{INTEGRAL} sources, IGR 
J18249--3243, IGR J19443+2117 and IGR J22292+6647, localized their soft X-ray counterparts and searched for 
associations with objects in the radio band. We also combined X-/gamma-ray data, as well as all the 
available radio, infrared and optical/UV information, in order to provide a broad-band spectral 
characterization of each source and investigate its nature.}
   {In all three cases, XRT observations provided a firm localization of the X-ray counterpart and 
information on its optical/UV, infrared and radio associations. All three sources are found to be bright 
and repeatedly observed radio objects, although poorly studied. The X-/gamma-ray spectrum of each source 
is well described by power laws with photon indices typical of AGN; only IGR J19443+2117 may have 
absorption in excess of the Galactic value, while IGR J22292+6647 is certainly variable at X-ray energies. 
IGR J18249--3243 has a complex radio morphology and a steep radio spectrum; the other two sources show 
flatter radio spectra and a more compact morphology. Overall, their radio, optical/UV and infrared 
characteristics, as well as their X-/gamma-ray properties, point to an AGN classification for
all three objects.} 
{}


   \maketitle

\begin{table*}
\begin{center}
\footnotesize
\caption{Log of \emph{Swift} XRT observations and source position.}
\label{Tab1}
\begin{tabular}{lcccccc}
\hline
\hline
Source & Obs. date & Exposure time$^{a}$ & Count rate (0.3--10 keV) & \multicolumn{3}{c}{Position} \\
    &             &     &    &    R.A.   & Dec.   &  Error     \\
       &          &      (s)     & (counts s$^{-1}$) & (J2000.0)   & (J2000.0)  & (arcsec)   \\
\hline
\hline
IGR J18249--3243  & Apr 08, 2007 &  4610  &  $0.111\pm0.005$  & 18 24 56.11 & --32 42 58.9 & 3.7 \\
\hline
IGR J19443+2117   & Oct 10, 2006 &  11078 &  $0.270\pm0.005$  & 19 43 56.20 & +21 18 22.9 & 3.5 \\
\hline
IGR J22292+6647   & May 19, 2007 &  9617       &  $0.075\pm0.003$   & 22 29 13.50 & +66 46 51.8 & 3.6  \\
                  & May 30, 2007 &  1703       &  $0.139\pm0.001$   \\
                  & Jun 16, 2007 &  1377       &  $0.134\pm0.001$   \\
                  & Jul 10, 2007 &  1656       &  $0.147\pm0.001$   \\
                  & Jul 14, 2007 &  385$^{b}$  &  $0.092\pm0.015$   \\
                  & Jul 20, 2007 &  3145       &  $0.075\pm0.005$   \\
                  & Aug 01, 2007 &  2104       &  $0.085\pm0.006$   \\
                  & Aug 02, 2007 &  467$^{b}$  &  $0.096\pm0.015$   \\
\hline
\hline
\end{tabular}
\end{center}
$^{a}$ Total on-source exposure time;\\
$^{b}$ This observation is not included in the fit procedure because of the too low exposure.
\end{table*}

\begin{table*}
\begin{center}
\footnotesize
\caption{Best-fit parameters as derived by fitting the XRT and IBIS data.}
\label{Tab2}
\begin{tabular}{lcccccccc}
\hline
\hline
Source & Energy band & $N_{\rm H(Gal)}^{a}$ & $N_{\rm H}^{a}$ & $\Gamma$&$C_{\rm calib}^{b}$ & 
$\chi^2/\nu$ &
$F_{\rm (2-10~keV)}^{c}$ & $F_{\rm (20-100~keV)}^{c}$ \\
       &   (keV) &  &    &  &  &  &   &      \\
\hline
\hline
IGR J18249--3243  & 0.5--100 & 0.118  & --  & $1.54\pm0.12$ & $0.52^{+0.35}_{-0.21}$ & 32.7/31  &
$0.52\pm0.03$  &  $0.79\pm0.15$  \\
\hline
IGR J19443+2117   & 0.9--100 & 0.836  &  $0.54^{+0.12}_{-0.13}$ & $2.04\pm0.12$ & $0.60^{+0.37}_{-0.28}$ 
& 84.3/93 & $1.83\pm0.04$  &  $1.12\pm0.22$  \\
\hline
IGR J22292+6647 (low + high) & 1--100 & 0.481 &  --  & $1.60\pm0.10$ & $0.76^{+0.45}_{-0.32}$$^{d}$&
120.5/105  &  $0.67\pm0.05$$^{e}$  & $1.10\pm0.02$  \\
\hline
\hline
\end{tabular}
\end{center}
$^{a}$ In units of $10^{22}$ cm$^{-2}$ (from Kalberla et al. 2005);\\
$^{b}$ IBIS/XRT cross-calibration constant;\\
$^{c}$ In units of $10^{-11}$ erg cm$^{-2}$;\\
$^{d}$ The simultaneous fit of the low and high XRT states, performed by normalizing to the low 
state, provides a cross-calibration constant for the high state of $1.42^{+0.16}_{-0.14}$, thus 
confirming variability during XRT observations;\\
$^{e}$ In this case we report the average value found for the flux.

\end{table*}

\section{Introduction}

A key strategic objective of the \emph{INTEGRAL} mission is a survey of the sky at high energies ($>$ 20 
keV), the domain where fundamental changes from primarily thermal to non-thermal sources/phenomena are 
expected, where the effects of absorption are drastically reduced and where most of the extreme 
astrophysical behaviour is taking place. To survey the high energy sky, \emph{INTEGRAL} makes use of the 
unique imaging capability of the IBIS instrument, which allows the detection of sources at the mCrab flux 
level with an angular resolution of 12$^{\prime}$ and a point source location accuracy of typically 
1--3$^{\prime}$ within a large ($29\times29$ degrees) field of view. 

So far, several surveys produced 
from data collected by IBIS have been reported in the literature, the most complete being that of Bird et 
al. (2007), which lists more than 400 sources of various nature (Galactic and extragalactic) and class. 
However, many ($\sim$25\%) of these new \emph{INTEGRAL} sources have no obvious counterpart in 
other wavebands and cannot be firmly classified; their classification is a primary objective of the 
survey work but it is made difficult by the arcmin size of the IBIS error boxes. Improved 
arcsec localization is therefore necessary to pinpoint the X-ray/optical counterpart and through 
spectroscopic observations assess its nature/class (e.g. Masetti et al. 2008 and references 
therein)\footnote{The catalogue of \emph{INTEGRAL} sources identified through optical and near-infrared 
spectroscopy is available at http://www.iasfbo.inaf.it/extras/IGR/main.html.}. 
Furthermore, many of these IBIS detections are newly discovered X/gamma-ray emitters and lack spectral 
information in the 2--10 keV band. Data in this waveband are fundamental to estimate the intrinsic column 
density via a measurement of the photoelectric cut-off and to have an indication of the source broad-band 
spectral shape. 

Cross correlations with catalogues in other wavebands, \emph{in primis} radio, can also 
be employed as a useful tool with which to facilitate the identification process and to identify peculiar and 
interesting objects. There are many valid reasons to use radio catalogues for counterpart searches: many 
high energy emitting objects are also powerful radio sources, for example active galactic nuclei (AGNs), 
pulsars, microquasars and cataclysmic variables (CVs); radio surveys are very sensitive and positionally 
accurate; the radio band does not suffer from the absorption which may be a limitation in other wavebands 
particularly for sources in the Galactic plane and center. This is quite important in the case where an 
optical counterpart is not found, for example due to heavy absorption or when many objects are present 
within the restricted X-ray error box: radio studies on top of the X-/gamma-ray ones, could help 
establish the nature of some of these newly discovered \emph{INTEGRAL} sources. 

In this work we provide 
radio, optical/UV, infrared and X-ray information on three objects extracted from the third IBIS catalogue 
(IGR J18249--3243, IGR J19443+2117 and IGR J22292+6647) and argue in favour of an AGN
nature for all three of them.

\section{X-/gamma-ray data}

All three sources are reported in the third \emph{INTEGRAL}/IBIS survey (Bird et al. 2007) where 
position and relative uncertainty, as well as 20--40 and 40--100 keV fluxes can be found. All three 
are located close to the Galactic plane since IGR J18249--3243, IGR J19443+2117 and IGR J22292+6647 
have Galactic latitudes of --9.2, --1.3 and +7.7 degrees, respectively. For this work, we use IBIS 
spectral data obtained following a non standard method, but using the same data set of Bird et al. 
(2007). The maximum source detection in a single IBIS pointing for all three objects is quite low; 
consequently, it is not possible to extract a good spectrum from any individual pointing. Instead, for 
each source all available pointings were processed with the OSA v. 5.1 software to produce images in 
various energy channels spanning the 20--100 keV range; channels over this band were logarithmically 
spaced to evenly distribute the counts across the channels. The weighted average flux of the source in 
each energy channel was then calculated from the individual OSA fluxes and variance images and used to 
construct a standard spectral \textit{pha} file. An appropriate rebinned \textit{rmf} file was 
produced from the standard IBIS spectral response file to match the chosen energy channels. Here and 
in the following, spectral analysis was performed with XSPEC v. 11.2.3 (Arnaud 1996) and errors are 
quoted at 90\% confidence level for one parameter of interest ($\Delta\chi^{2}=2.71$).

\begin{figure} 
\centering
\includegraphics[width=0.65\linewidth,angle=-90]{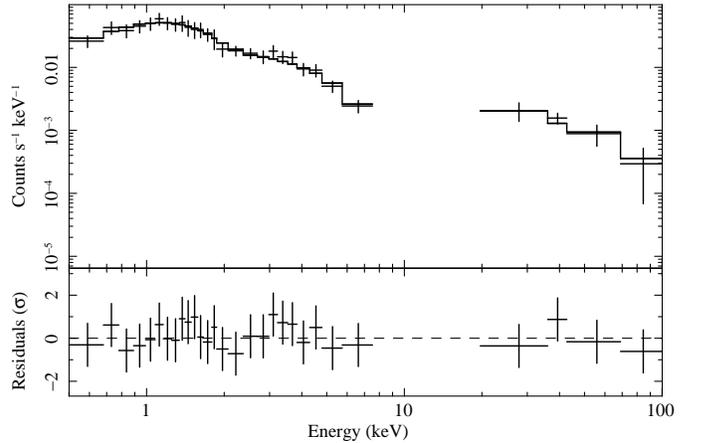}
\caption{XRT/IBIS broad-band spectrum of IGR J18249--3243 fitted with a simple power law (upper 
panel); residuals to this model are in units of $\sigma$ (lower panel).} 
\label{fig1}
\end{figure}

\begin{figure} 
\centering
\includegraphics[width=0.65\linewidth,angle=-90]{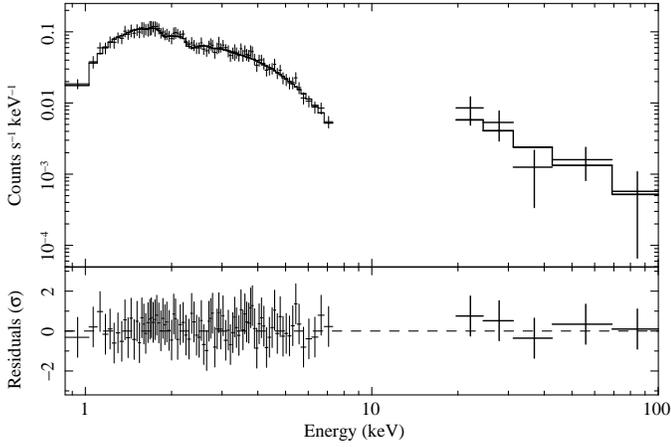}
\caption{XRT/IBIS broad-band spectrum of IGR J19443+2117 fitted with an absorbed power law; 
residuals to this model are in units of $\sigma$ (lower panel).} 
\label{fig2}
\end{figure}

\begin{figure}
\centering
\includegraphics[width=0.65\linewidth,angle=-90]{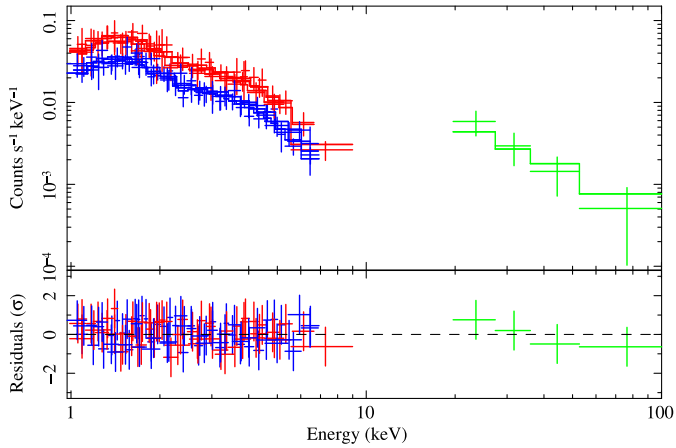}
\caption{XRT/IBIS broad-band spectrum of IGR J22292+6647 in low state (blue) and high state
(red) fitted with a simple power law; residuals to this model in units of $\sigma$.}
\label{fig3}
\end{figure}   

Due to the low signal to noise ratio of these detections, a simple power 
law has been employed to fit the IBIS data, providing $\Gamma$ values in the range $\sim$1.5--2.0.

For each source, we also have X-ray observations acquired with the XRT (X-ray Telescope, 0.2--10 keV, 
Burrows et al. 2005) on board the \emph{Swift} satellite (Gehrels et al. 2004), which allowed us to better 
determine the position of the IBIS counterpart. XRT data reduction was performed using the XRTDAS standard 
data pipeline package ({\sc xrtpipeline} v. 0.11.6), in order to produce screened event files. All data 
were extracted only in the Photon Counting (PC) mode (Hill et al. 2004), adopting the standard grade 
filtering (0--12 for PC) according to the XRT nomenclature. Images have been extracted in the 0.3--10 keV 
band and searched for significant excesses falling within or around the \emph{INTEGRAL}/IBIS 90\% 
confidence circle as reported in Bird et al. (2007); in all cases a single bright X-ray source was detected 
inside this uncertainty circle or just at its border. The log of all X-ray observations is given in 
Table~\ref{Tab1}. The 90\% uncertainty was obtained using the task {\sc xrtcentroid} v. 0.2.7. The XRT 
positions of IGR J18249--4243 and IGR J19443+2117 are compatible with the location of two ROSAT Bright 
sources (1RXS J182456.2--324329 and 1RXS J194356.1+211731, respectively). IGR J22292+6647 is instead likely 
associated with a ROSAT Faint object (1RXS J222915.7+664704) and coincides with a \emph{XMM-Newton} Slew 
Survey object (XMMSL1 J222914.4+664653); this last measurement also provides a 0.2--12 keV flux of 
$3\times10^{-12}$ erg cm$^{-2}$ s$^{-1}$.

Events for spectral analysis were extracted within a circular region of radius 20$^{\prime 
\prime}$, centered on the source position, which encloses about 90\% of the PSF at 1.5 keV (see Moretti et 
al. 2004). The background was taken from various source-free regions close to the X-ray source of 
interest, using circular regions with different radii in order to ensure an evenly sampled background. In 
all cases, the spectra were extracted from the corresponding event files using the {\sc XSELECT v. 2.4} 
software and binned using {\sc grppha}, so that the $\chi^{2}$ statistic could be 
applied. We used version v. 009 of the response matrices and created individual ancillary response 
files \textit{arf} using the task {\sc xrtmkarf v. 0.5.6}. 

Only IGR J22292+6647 has more than one pointing; in 
this case we performed the spectral analysis of those observations having sufficient statistics (6 out of 
8 pointings). We found changes in fluxes ($\sim$50$\%$) but not in shape; we therefore combined 
measurements with similar count rates to provide a low (LS) and a high (HS) X-ray state for 
comparison with the average IBIS spectrum. 

Next, we analyse the broad-band data of each source over the 0.5--100 keV range. The results of the 
spectral analysis are listed in Table~\ref{Tab2}. The best-fits are given by simple power law absorbed by 
Galactic absorption and, when required, also by adding an intrinsic absorption; we also introduce in the 
fitting procedure a cross-calibration constant ($C_{\rm calib}$) to account for a possible mismatch 
between XRT and IBIS data as well as for source flux variations. Only in the case of IGR J19443+2117 do we 
find evidence for absorption in excess of the Galactic value. The addition of this component strongly 
improves the $\chi^2$ value, being significant at more than 99.99\% confidence level ($\Delta\chi^{2}=73$ 
for one degree of freedom), according to the $F$-test, and gives a column density of 
$\sim$$8\times10^{21}$ cm$^{-2}$. Photon indices range again from 1.5 to 1.6, i.e. similar to those 
typically found in AGNs.

In all sources, the fact that the cross-calibration is different from 1 suggests some flux 
variability. This is more evident in IGR J22292+6647 due to the larger data set available; indeed in this 
source \emph{XMM-Newton} detected a lower X-ray flux (especially in view of the larger band used) than 
those reported in Table~\ref{Tab2}. Figures 1, 2 and 3 show the broad-band X-ray data of each source and, 
for IGR J22292+6647, also the low and high XRT states (in red and blue in Figure~\ref{fig3}).

\begin{figure*} 
\includegraphics[width=0.36\linewidth]{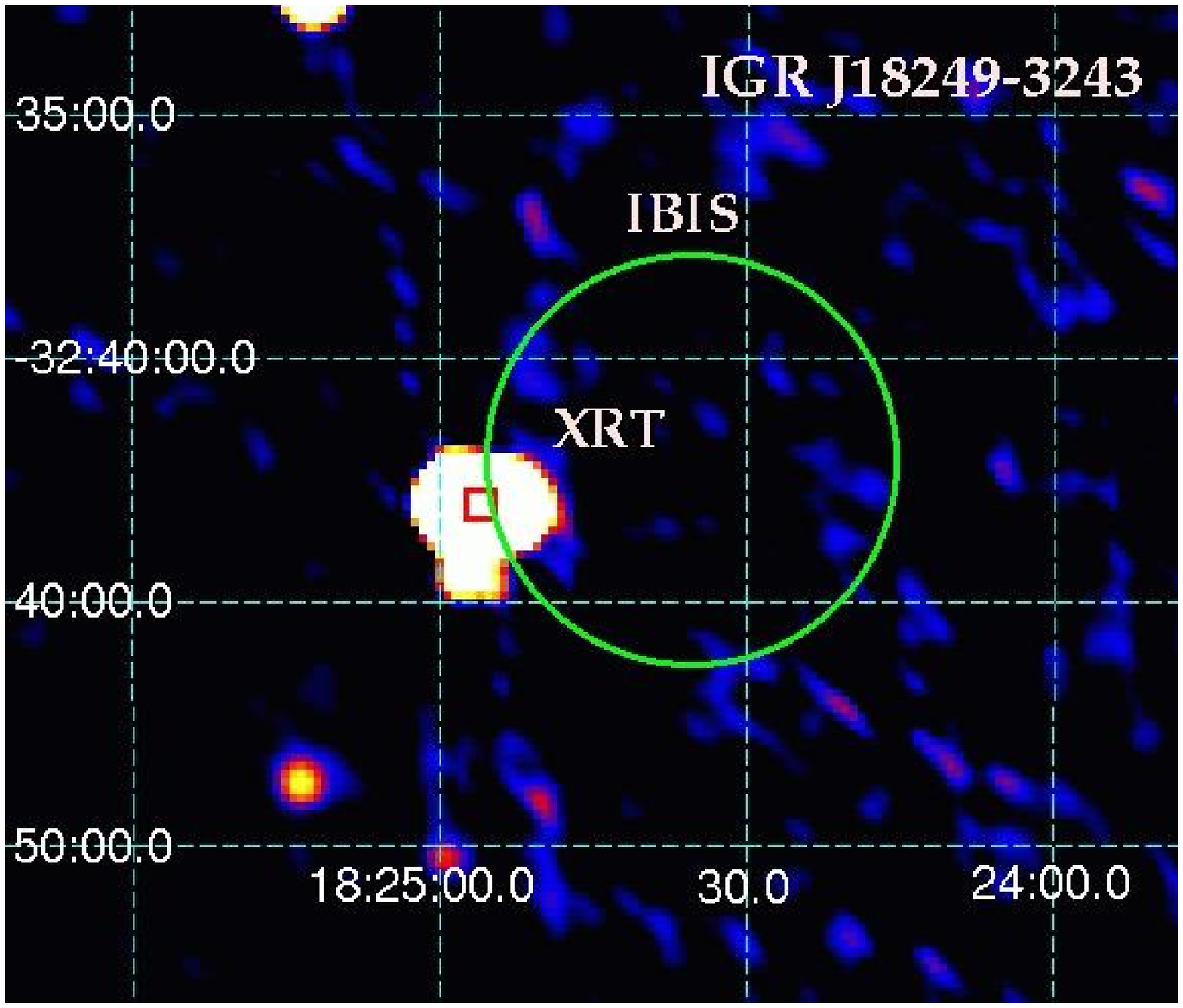}
\includegraphics[width=0.45\linewidth]{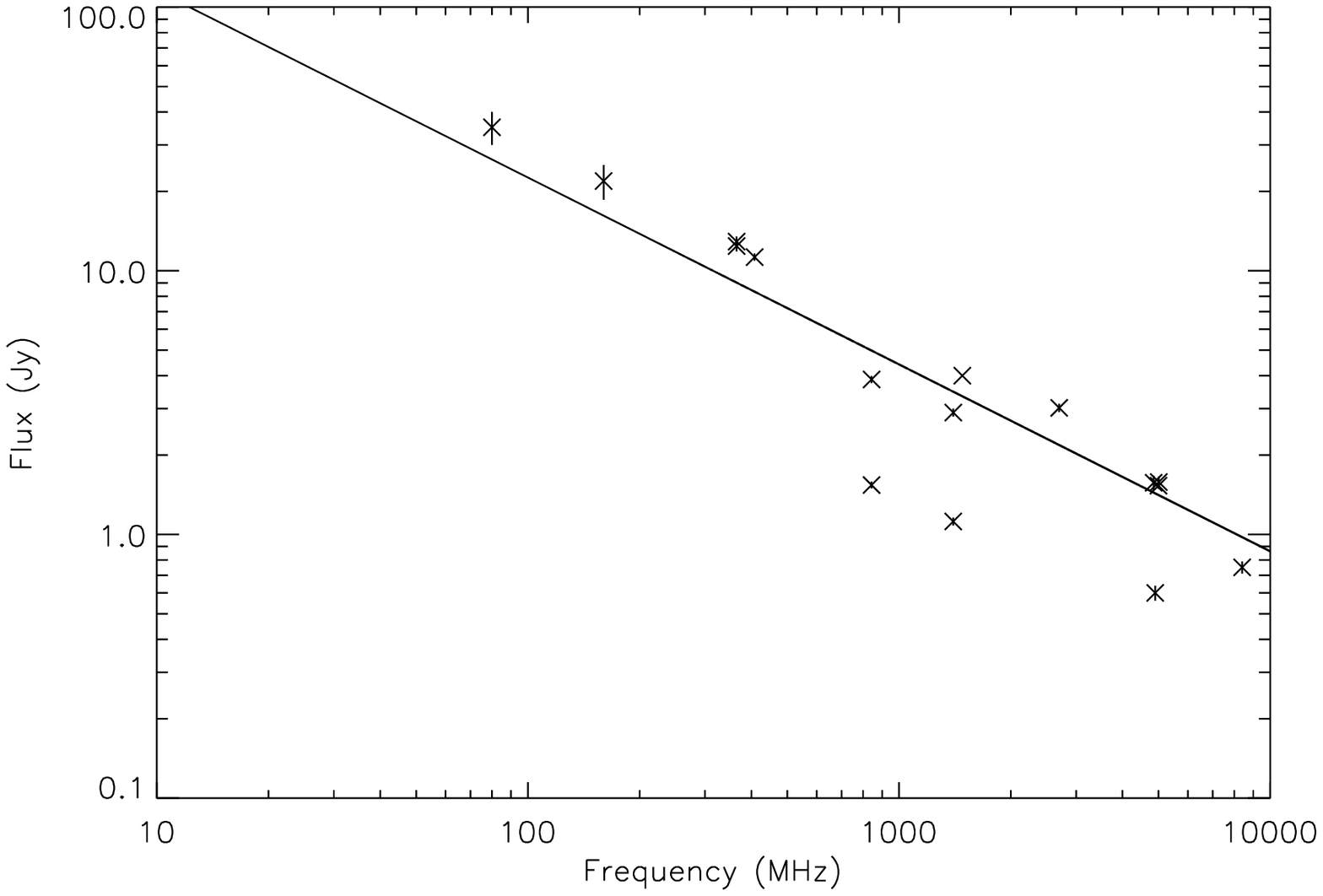}
\includegraphics[width=0.36\linewidth]{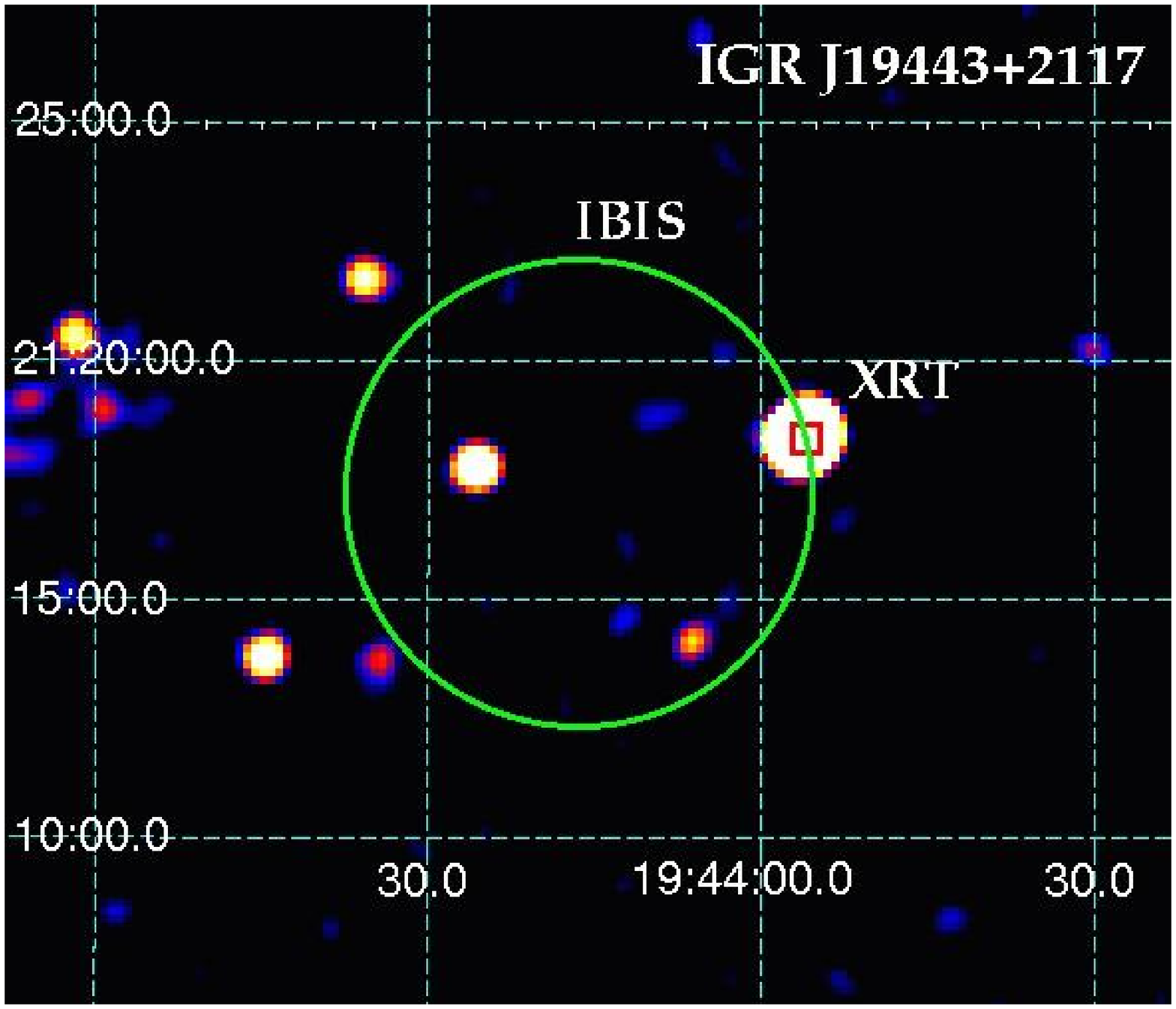}
\includegraphics[width=0.45\linewidth]{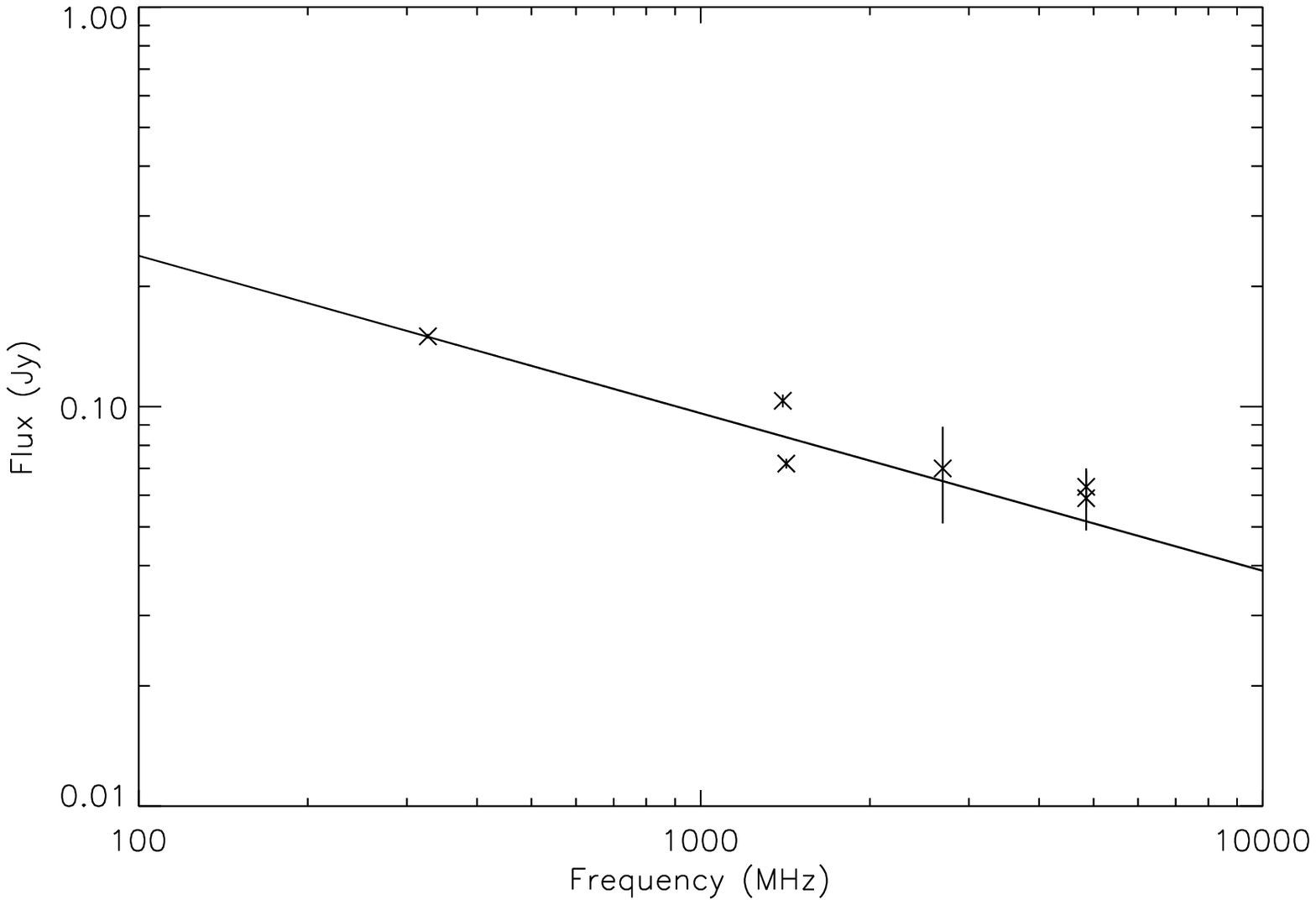}
\includegraphics[width=0.36\linewidth]{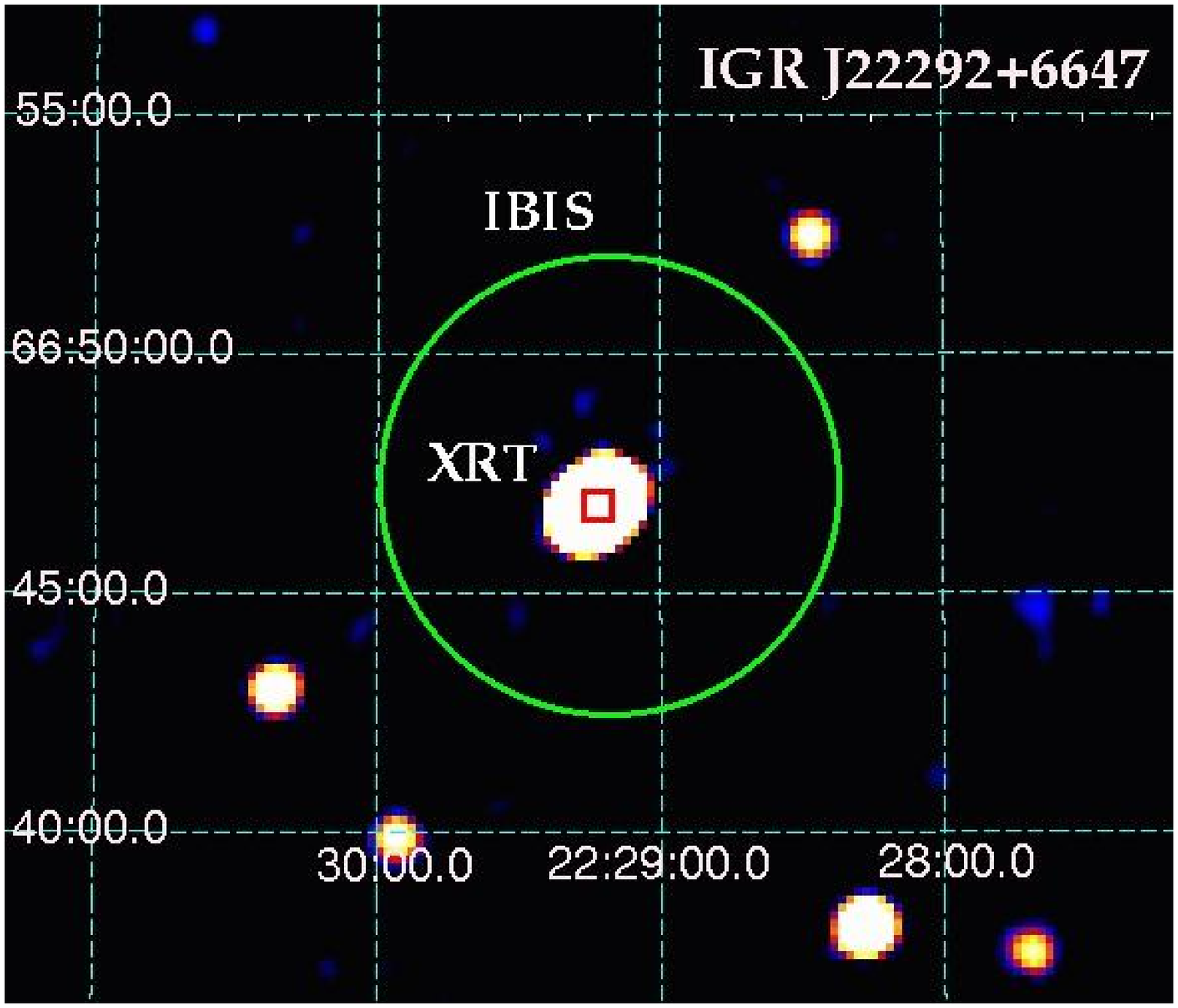}
\hspace{3.5cm}
\includegraphics[width=0.45\linewidth]{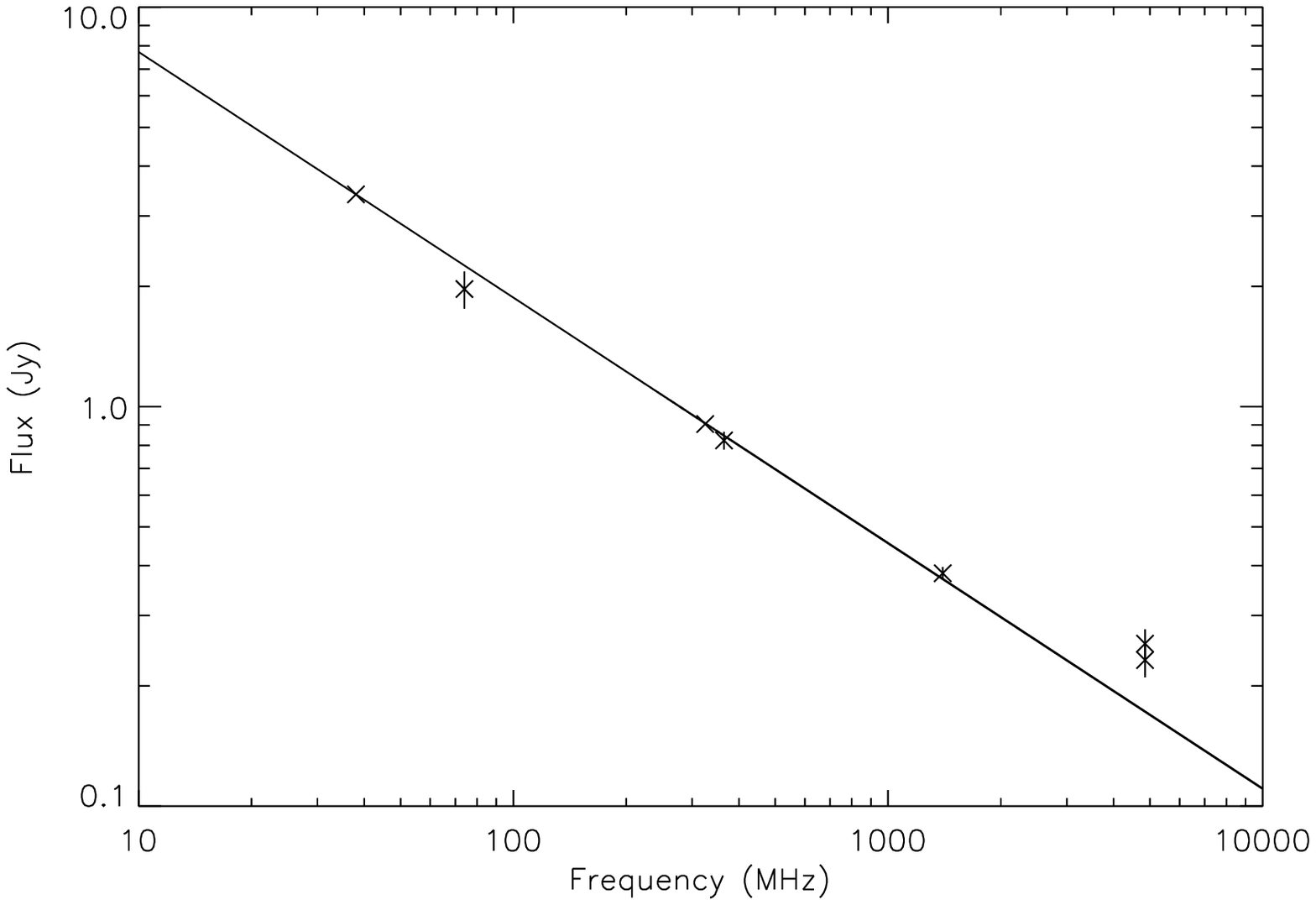}
\caption{NVSS images and radio spectra fitted with $Log(S(\nu)) = a \times Log(\nu) + b$ 
of IGR J18249--3243 (upper left and right panels), IGR J19443+2117 (central left and right panels) and 
IGR J22292+6647 (lower left and right panels). The green circle describes the IBIS uncertainty, 
while the XRT position is given by the red box.} 
\label{fig4}
\end{figure*}

\section{Radio data}

Historically AGNs were discovered by radio observations, i.e. the radio selection is often a way to 
recognize active galaxies, except at lower luminosities where star-formation in galaxies can provide radio 
emission. Therefore, for bright objects mere detection in radio provides support for the presence of an 
active galaxy. Contamination from Galactic sources may come from pulsars, microquasars and CVs. 
In some cases, the radio morphology can help in discriminating between the above 
possibilities since radio sources showing complex structures (double lobe morphology or hint for jets) are 
often associated with AGN; furthermore, radio spectra ($S \propto \nu^{\alpha}$) of AGNs are well 
recognised as being flat ($\alpha = -0.3/-0.5$) in compact sources and steep ($\alpha=-0.8/1.0$) in 
complex objects. Pulsars often have undetected or very dim optical/infrared counterparts, while 
microquasars are very rare objects. Of the CVs only magnetic systems may be associated to radio 
emission. So, while mere radio detection does not 
imply identification with an AGN, the combination of high energy X-/gamma-ray emission together with 
detection in the optical, UV and infrared bands plus complex radio morphology and/or an AGN radio spectrum 
provide strong support for the AGN nature of an unclassified \emph{INTEGRAL} source.

\begin{table}
\caption{Summary of the UVOT observations of the three sources. The
magnitudes are not corrected for Galactic reddening (Schlegel et al. 1998).}
\label{Tab3}
\begin{flushleft}
\begin{tabular}{lccc}
\hline
\hline
\noalign{\smallskip}
Source & Filter & $T_{\rm exp}$ & Magnitude \\
\noalign{\smallskip}\hline\noalign{\smallskip}
\hline
IGR J18249--3243  & UVW2  &    4727   &  17.53$\pm$0.02 \\
\hline
IGR J19443+2117   &  V  &  907 & $>$20.3 (3$\sigma$ ul) \\
                  &  B  &  907 & $>$21.2 (3$\sigma$ ul) \\
                  &  U  &  907 & $>$20.8 (3$\sigma$ ul)  \\
                  & UVW1  &    1826   & $>$21.2 (3$\sigma$ ul)  \\
                  & UVM2  &    2518   & $>$21.5 (3$\sigma$ ul)  \\
                  & UVW2  &    1417   & $>$21.3 (3$\sigma$ ul)  \\
\hline
IGR J22292+6647 & UVW1  &   11516   & $>$21.83 (3$\sigma$ ul)  \\
     & UVM2  &     461   & $>$20.41 (3$\sigma$ ul)  \\
     & UVW2  &    5179   & $>$21.64 (3$\sigma$ ul)  \\       
\hline
\noalign{\smallskip}\hline\noalign{\smallskip}
\end{tabular}
\end{flushleft}
\end{table}

For this study, radio images have been taken primarily from the NVSS (NRAO VLA Sky Survey, Condon et al. 
1998) at 1.4 GHz. All images have 45 arcsecond FWHM angular resolution and nearly uniform sensitivity. The 
rms uncertainties in right ascension and declination vary from about 1 arcsecond for the sources stronger 
than 15 mJy to 7 arcseconds at the survey limit of 2.5 mJy. The left panels of Figure~\ref{fig4} show the 
NVSS image cut-outs with overimposed IBIS error circle and XRT position. All three objects have, within 
relative uncertainties, a radio counterpart coincident with the X-ray source. IGR J18249--3243 is a 
complex source in the NVSS map with at least three components reported (NVSS J182457--324259 with 
$2.898\pm0.101$ Jy flux, NVSS J182454--324257 with $1.116\pm0.038$ Jy flux and NVSS J182457--324419 with a 
$0.0226\pm0.0011$ Jy flux); similarly, two very close components are found in the Molonglo Galactic Plane 
Survey 2nd Epoch (MGPS-2) Compact Source Catalogue (Murphy et al. 2007). The source has been extensively 
observed at radio frequencies and appears in many radio catalogues, although its nature has been poorly 
studied so far. It is also reported as an asymmetric double source in NED (see also the Texas Survey of 
Radio Sources at 365 MHz, Douglas et al. 1996). To provide more information on this and the other two 
sources, we have used SpecFind (Vollmer et al. 2005) and CATS (the on-line Astrophysical CATalogs support 
System, at http://cats.sao.ru/; see also Verkhodanov et al. 1997); both are tools used to cross-identify 
radio sources in various catalogues on the basis of self-consistent spectral index as well as position. 
This allows us to combine data at different frequencies and to estimate the source radio spectrum as 
$Log(S(\nu)) = a \times Log(\nu)
+ b$, where $S$ in expressed in Jy and $\nu$ in MHz. For IGR J18249--3243 $a$ is $-0.71\pm0.01$ and 
$b$ is $2.77\pm0.03$; the 
corresponding source spectrum is displayed in the upper right panel of Figure~\ref{fig4}. 

The structure of IGR J19443+2117 is less complex, with only one component reported as NVSS J194356+211826 
with a 1.4 GHz 
flux of $0.103\pm0.004$ Jy. Also this source appears in various radio catalogues but again its nature and 
class are not yet defined. Its radio spectrum is plotted in the central right panel of Figure~\ref{fig4} 
and the spectral parameters in this case are $a= -0.394\pm0.016$ and $b=0.165\pm0.044$. 

IGR J22292+6647 has also one component in the NVSS (NVSS J222913+664654) with a 1.4 GHz flux of 
$0.381\pm0.014$ Jy. Similarly to the other two objects, it is reported in many radio catalogues including 
the recent VLA Low-Frequency Sky Survey (VLSS, Cohen et al. 2007). The spectral parameters are $a= 
-0.615\pm0.002$ and $b=1.503\pm0.003$ and the source radio spectrum is shown in the lower panel of 
Figure~\ref{fig4}; also IGR J22292+6647 is reported as an asymmetric double in NED (see also the Texas 
Survey of Radio Sources at 365 MHz).

\begin{table}
\caption{Summary of the diagnostic parameters.}
\label{Tab4}
\begin{flushleft}
\begin{tabular}{lcccc}
\hline
\hline
\noalign{\smallskip}
Source & $\alpha_{\rm ox}$ & $\alpha_{\rm ro}$ & $\alpha_{\rm rx}$  & log($R_{\rm X}$) \\
\noalign{\smallskip}\hline\noalign{\smallskip}
\hline
IGR J18249--3243$^{a}$  & 1.15 & 0.75 & 0.86 & --1.82 \\
\hline
IGR J19443+2117   &  $>$1.38 & $<$0.26 & 0.63 & --3.74 \\
\hline
IGR J22292+6647 & $>$1.63  & $<$0.32   & 0.75 &  --2.74 \\
\hline
\noalign{\smallskip}\hline\noalign{\smallskip}
\end{tabular}
\end{flushleft}
$^{a}$ For this source $\alpha_{\rm ox}$ and $\alpha_{\rm ro}$ have been estimated 
by extrapolating the UVW2 flux to 2500 $\AA$ 
and using an average UV slope of $\alpha_{\rm uv} =0.4$ (Vanden Berk et al. 2001).
\end{table}

\section{Ultraviolet, optical and infrared data} 

All three sources were observed by the \emph{Swift} UV/Optical telescope (UVOT, Roming et al. 2005). 
Information regarding UVOT observations are reported in Table~\ref{Tab3}. The 
data were reduced in the standard way by coadding the exposures in each filter by the UVOT task  
\emph{uvotimsum}. Source counts were selected with the standard 5$^{\prime \prime}$ radius for all UVOT 
filters according to the most recent UVOT photometry calibration as described by Poole et al. (2008).
Background photons were 
selected in a source-free region close-by with a radius of 20$^{\prime \prime}$. Magnitudes 
were measured with the UVOT tool \emph{uvotsource}.

Thanks to the spectral information available over a broad range of frequencies, we can determine 
(see Table~\ref{Tab4}) the 
optical-to-X-ray ($\alpha_{\rm ox} = log(S_{\rm 2~keV}/S_{\rm 2500~\AA})/2.605$), radio-to-optical
($\alpha_{\rm ro} = log(S_{\rm 5~GHz}/S_{\rm 2500~\AA})/5.38$), 
radio-to-X-ray ($\alpha_{\rm rx} = log(S_{\rm 5~GHz}/S_{\rm 2~keV})/7.99$) indices and the 
X-ray radio-loudness parameter $R_{\rm X} = S_{\rm 4.85~GHz}/S_{\rm 2-10~keV}$ (see Stocke et al. 
1991; Terashima \& Wilson 2003). 
Following Panessa et al. (2007) log($R_{\rm X}) > -2.755$ can be used as a boundary to determine if an AGN 
is radio loud.

Within the XRT uncertainty circle of IGR J18249--3243, there is a counterpart in the USNO--B1.0 
catalogue (Monet et al. 2003), with $R$ magnitude in the range 14.6--14.8, which is also listed in the 
2MASS 
survey (Skrutskie et al. 2006) with 
$J=13.456\pm0.026$, $H=13.007\pm0.029$ and $K=12.675\pm0.022$ and has a detection in 
$UVW2=17.53\pm0.02$ 
(see Table~\ref{Tab3}).
The Galactic color excess measured along 
the line of sight to this object is $E(B-V)=0.23$ (here and afterwards this value is estimated following 
Schlegel et al. 1998); this value translates to a Galactic column density of 
$0.13\times10^{22}$ cm$^{-2}$ (Predehl \& Schmitt 1995), i.e. compatible with the value quoted in 
Table~\ref{Tab2}. Recently, optical follow-up observations of IGR J18249--3243 have confirmed the AGN 
nature of this object despite difficulties in separating the active galaxy at $z=0.355$ from a foreground 
star (details will be found in Masetti et al. submitted to A\&A). It is therefore likely that the 
USNO--B1.0, 2MASS and UV magnitudes are contaminated by this star. However, we were able to provide an 
upper limit to the $R$ flux\footnote{Here and in the following the optical and infrared fluxes are 
corrected for Galactic absorption.} of $7.6\times 10^{-12}$ erg cm$^{-2}$ s$^{-1}$ (or 6.9 mJy), 
adopting the conversion factors of Fukugita et al. (1995). 
The diagnostic parameters of Table~\ref{Tab4} provide additional support 
for an active galaxy classification for IGR J18249--3243 and further suggest that it might be a 
radio loud AGN.

Within the XRT error box of IGR J19443+2117, we find a 2MASS counterpart located at
R.A.=19$^{\rm h}$43$^{\rm m}$56$^{\rm s}$.24 and Dec=+21$^{\circ}$ 18$^{\prime}$ 23$^{\prime 
\prime}$.4 (J2000.0), with magnitudes $J > 14.53$, $H > 13.36$ and $K = 13.980\pm 0.07$. 
Using the 2MASS conversion factors 
(Skrutskie et al. 2006), the infrared $K$ flux is $6.5\times10^{-13}$ erg cm$^{-2}$ s$^{-1}$ (or 3.8 mJy).
This infrared source has no counterpart in the USNO--B1.0 catalogue and \emph{Swift} UVOT images either 
(see Table~\ref{Tab3}),
implying a dim flux at optical/UV frequencies and difficulties in providing optical spectroscopy 
in the nearby future. Note also that it is outside the radio error box and may not even be the true 
counterpart.
The Galactic color excess measured in the source direction is $E(B-V)=2.59$, 
corresponding to a column density of $1.4\times 10^{22}$ cm$^{-2}$ (Predehl \& Schmitt 1995), 
significantly higher than the Galactic $N_{\rm H}$ value quoted in Table~\ref{Tab2}, but compatible with 
the total absorption measured in this source; this suggests some caution in considering IGR J19443+2117 as an 
intrinsically absorbed object. 
The diagnostic parameters listed in Table~\ref{Tab4} suggest that IGR J19443+2117 is an AGN
more likely of the radio quiet type.

\begin{figure} 
\centering
\includegraphics[width=0.80\linewidth]{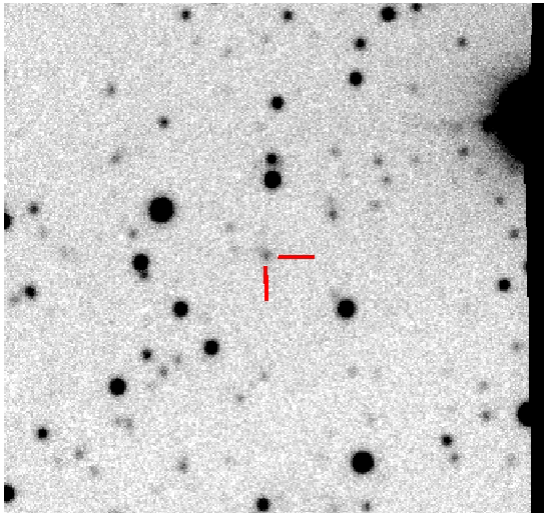}
\caption[]{Section of the 'Cassini' plus BFOSC $R$-band stacked
image (total exposure time: 3600 s) of the field of IGR J22292+6647
acquired on 7 November 2007.
The putative optical counterpart of this high-energy source is
indicated by the tick marks. In the image, of size 3$'$$\times$3$'$,
North is at top and East is to the left.}
\label{fig5}
\end{figure}  

The field of IGR J22292+6647 was observed from Loiano (Italy) with the 1.5m `G.D. Cassini' telescope plus 
BFOSC, equipped with a 1300$\times$1340 pixels EEV CCD, under non-optimal weather conditions (seeing: 
$\sim$3$''$). The BFOSC images have a plate scale of 0$\farcs$58/pix, and a useful field of 
13$\farcm$0$\times$12$\farcm$6. The imaging observations, performed in the Cousins $R$ band, started at 
17:41 UT of 7 November 2007 and were composed of two frames, both with exposure time of 1800 s. Imaging 
frames were corrected for bias and flat-field in the usual fashion, and were stacked together to improve 
the signal-to-noise ratio. Visual inspection of the stacked image (see Figure~\ref{fig5}) confirms the 
presence of an object, also faintly visible on the corresponding DSS--II--Red survey\footnote{available 
at {\tt http://archive.eso.org/dss/dss}} image, within the XRT error box. This source appears to be 
extended, therefore simple aperture photometry, within the 
MIDAS\footnote{\texttt{http://www.eso.org/projects/esomidas}} package, was used to measure its $R$-band 
magnitude. This was then calibrated using nearby USNO--A2.0 catalogue\footnote{available at \\ {\tt 
http://archive.eso.org/skycat/servers/usnoa/}} stars. Our final $R$-band photometry of the putative 
optical counterpart of IGR J22292+6647 yields a magnitude $R = 19.2\pm0.1$. This object is also listed in 
the 2MASS survey with magnitudes $J=16.602\pm0.143 $, $H = 15.707\pm0.163$ and $K = 14.683\pm0.126$. The 
location of IGR J22292+6647 in the $J-H$ versus $H-K$ color-color diagram is fully compatible with 
an AGN nature for this source (Spinoglio et al. 1995). The Galactic color excess measured along 
its line of sight is $E(B-V)=1.089$; this corresponds to a Galactic column density of $0.6\times10^{22}$ 
cm$^{-2}$ (Predehl \& Schmitt 1995), i.e. compatible with the value quoted in Table~\ref{Tab2}. 
Although the observed $R-K$ value is $\sim$4.5, i.e. suggestive of a very red object,
the corrected value is 2.1, fully compatible with typical AGN colors. 
The $R$ magnitude, estimated using the conversion factors of Fukugita et al. (1995), gives a flux,
corrected for Galactic absorption, of $8.9\times10^{-13}$ erg cm$^{-2}$ s$^{-1}$ (or 0.8 mJy).
The diagnostic parameters (see Table~\ref{Tab4}) support an AGN
nature for IGR J22292+6647 too and further suggests it might also be a borderline object in between the 
radio loud and radio quiet regimes.

\section{Conclusions}

We have found the X-ray (and consequently optical and/or UV/infrared) counterparts of three 
\emph{INTEGRAL} unidentified/unclassified sources. The restricted X-ray positions have allowed the 
association of all three with bright radio emitters often reported in various catalogues but not yet 
properly studied.

IGR J18249--3243 is the brightest and most complex in radio of the three analysed here and also the 
farthest 
away from the Galactic plane; both morphology and spectrum at radio frequencies are compatible with 
an AGN nature. The X-ray spectrum is flat but still within the range observed in AGN. 
The diagnostic parameters listed in Table~\ref{Tab4}
support the AGN classification and also indicate that the source is radio loud.
Recent optical spectroscopy confirms our finding locating the source at $z=0.355$ (Masetti et al. 
submitted to A\&A).

The joint XRT and IBIS spectrum of IGR 
J19443+2117 is also typical of an AGN. The source is moderately bright at radio frequencies and 
shows a flat spectrum. It is very dim at optical/UV frequencies 
so that its classification will not be easy; as the source is the closest to the Galactic plane and 
with the least information available, its classification remains uncertain and 
therefore we propose it as an AGN candidate, probably radio quiet.

IGR J22292+6647 is the source best studied at high energies: 
its LS and HS spectra are well fitted with simple power laws having, within the uncertainties, the same 
photon index but variable fluxes. At radio frequencies the source is quite bright, possibly shows a 
complex morphology (asymmetric double?) and has a flat spectrum; all these, combined with the infrared
photometry and the diagnostic parameters shown in Table~\ref{Tab4}, are indicative of an AGN nature; 
it is possible that this source is a borderline object in between the radio loud and radio 
quiet regimes.

Since optical follow-up spectroscopy of IGR J19443+2117 and IGR J22292+6647 will not be 
straightforward, 
we encouraged radio measurements of both objects to confirm our proposed classification.

\begin{acknowledgements} 

We thank Silvia Galleti and Roberto Gualandi for Service Mode observations at the `Cassini' telescope in 
Loiano. We also thank Marco Salvati for useful remarks
which helped us to improve the paper.
This research has made use of data obtained from the High Energy Astrophysics Science Archive 
Research Center (HEASARC), provided by NASA's Goddard Space Flight Center, the NASA/IPAC Extragalactic 
Database (NED), the USNO--B1.0 and 2MASS catalogues and the CATS database (Astrophysical CATalogs support 
System). The authors acknowledge the ASI financial support via ASI--INAF grants I/088/06/0 and 
I/008/07/0. \emph{Swift} at PSU is supported by NASA contract NAS5--00136. This research was also 
supported by NASA contract NNX07AH67G (D.G.). We also acknowledge the use of public data from the 
\emph{Swift} data archive. 

\end{acknowledgements}

\end{document}